\documentclass[twocolumn,aps,pra,superscriptaddress,nofootinbib]{revtex4}

\usepackage{amsmath}
\usepackage{amssymb}
\usepackage[usenames,dvipsnames]{color} 
\usepackage{epsf}
\usepackage{epsfig}
\usepackage{graphicx}

\newcommand\Id{\leavevmode\hbox{\small1\normalsize\kern-.33em1}}
\newcommand{\half}{\frac{1}{2}}

\newcommand{\carb}{$^{13}$C }
\newcommand{\ket}[1]{\left\vert{#1}\right\rangle}

\newcommand{\ku}{\vert{0}\rangle}
\newcommand{\kd}{\vert{1}\rangle}

\newcommand{\bu}{\langle{0}\vert}
\newcommand{\bd}{\langle{1}\vert}

\newcommand{\ave}[1]{\left\langle #1\right\rangle}
\newcommand{\ham}{{\mathcal{H}}}

\newcommand{\tr}[1]{\textrm{Tr}\left\{{#1}\right\}}

\newcommand{\pdev}[2]{\frac{\partial #1}{\partial #2}}

\definecolor{DarkBlue}{rgb}{0,0,.4}

\DeclareMathSymbol{\vartheta}{\mathalpha}{letters}{"12}
\DeclareMathSymbol{\theta}{\mathalpha}{letters}{"23}
\DeclareMathSymbol{\phi}{\mathalpha}{letters}{"27}
\DeclareMathSymbol{\varphi}{\mathalpha}{letters}{"1E}

\definecolor{LinkColor}{rgb}{0,0,.5}
\usepackage[colorlinks=true,linkcolor=BrickRed,citecolor=LinkColor,urlcolor=LinkColor]{hyperref}
\renewcommand{\emph}{\textit}

\begin{document}

\title{Environment Assisted  Metrology with Spin Qubits}
\author{P. Cappellaro}\email{pcappell@mit.edu}\affiliation{Nuclear Science and Engineering Dept., Massachusetts Institute of Technology, Cambridge MA 02139 USA}
\author{G. Goldstein}
\affiliation{Department of Physics, Harvard University, Cambridge MA 02138 USA}
\author{J. S. Hodges}\altaffiliation{Current address: Quantum Information Science Group, MITRE Corp. 260 Industrial Way West, Eatontown, NJ 07724, USA}
\affiliation{Nuclear Science and Engineering Dept., Massachusetts Institute of Technology, Cambridge MA 02139 USA}
\affiliation{Department of Physics, Harvard University, Cambridge MA 02138 USA}
\author{L. Jiang}
\affiliation{Institute for Quantum Information, California Institute of Technology, Pasadena CA 91125 USA}
\author{J. R. Maze}
\affiliation{Faculty of Physics, Pontificia Universidad Catolica de Chile, Santiago 7820436, Chile}
\author{A. S. S{\o}rensen}\affiliation{QUANTOP, Niels Bohr Institute, Copenhagen University, DK
2100, Denmark}
\author{M. D. Lukin}\affiliation{Department of Physics, Harvard University, Cambridge MA 02138 USA}

\begin{abstract}
We investigate the sensitivity of a recently proposed method for precision measurement  
[Phys. Rev. Lett. {\bf 106}, 140502 (2011)], focusing on an implementation based on
solid-state spin systems. The scheme amplifies a quantum sensor response to weak external fields by exploiting its coupling to spin
impurities in the environment.
We analyze the limits to the sensitivity due to decoherence and propose dynamical decoupling schemes to increase the spin coherence time.
The sensitivity is also limited by the environment spin polarization; therefore we discuss strategies to polarize the environment spins and  
present a method to extend the scheme to the case of zero polarization.
The coherence time and polarization determine a figure of merit for the environment's ability to enhance the sensitivity compared to echo-based sensing schemes. 
This figure of merit 
 can be used to engineer optimized samples for high-sensitivity nanoscale magnetic sensing, such as diamond nanocrystals with controlled impurity density.
\end{abstract}
\maketitle

\section{Introduction}\label{Intro}

Quantum metrology seeks to achieve precision measurements with an accuracy beyond the limits imposed 
by the central limit theorem~\cite{Giovannetti11} (the standard quantum limit, SQL).
Although many proposals for achieving the quantum limits of sensitivity (as defined by the Heisenberg bounds) have been presented, they are often difficult to implement in practice. 
The main challenges arise from the deleterious effects of noise and decoherence on the (entangled) states required for quantum metrology  and from the unavailability of the Hamiltonians and measurement strategies needed to create and readout these entangled states.

We recently introduced a scheme~\cite{Goldstein11} that aims at overcoming these two challenges. We proposed to use the environment of the sensor as an additional resource for metrology and we showed how to achieve the desired interaction Hamiltonian using coherent control techniques. 
In this paper we focus on one possible implementation of this \textit{environment assisted metrology} (EAM) scheme -- a spin sensor embedded in a bath of other spins -- in order to derive more detailed results on the sensitivity achievable. In addition, we will analyze in depth the effects of decoherence and of finite polarization.

The paper is organized as follows. 
In Section~\ref{EAM} we present the EAM scheme: the control sequence that achieves it and the sensitivity gain in the idealized situation of no decoherence. 
This restriction is lifted in Section~\ref{Decoherence}, where we analyze the effects of decoherence, both analytically and with numerical simulations. We further provide strategies to reduce the effects of decoherence. In section~\ref{Sensitivity} we use these results to derive limits of the proposed EAM strategy and compare them to usual strategies that do not take advantage of the environment.  Since the sensitivity depends on the polarization of the spin environment, we propose in section~\ref{Polarization}  schemes for polarizing these ancillary spins and we  further extend the scheme to the case where no polarization is available. 
A second extension of the EAM method is presented in section~\ref{Spins}, where more general spin systems are studied.

\section{The environment assisted metrology scheme}\label{EAM}
We consider the metrology task of measuring a parameter $b$ via its interaction with a quantum probe. The task can be accomplished by using 
a Ramsey scheme, where a two level system is first prepared in a superposition of the two states, which then acquire a phase difference that is mapped onto the populations by a second pulse. An example of this scheme is magnetometry with solid-state spins~\cite{Taylor08}, where the probe interacts with the external magnetic field via a Hamiltonian $\ham\propto bS_z$, acquiring a phase $\propto bt$ during the interrogation time  $t$. Then, the bound to the sensitivity is set by the dephasing rate  that limits the time the probe can interact with the external field associated to the parameter to be measured. 

Coherent control techniques can be used to isolate the probe from its environment, thus increasing the coherence time. If the environment interacts as well with the external field to be measured -- as it is the case for a spin bath -- a different strategy is possible: 
in Ref. \cite{Goldstein11} we showed that the spin environment can be used as a resource in this case, by mapping the phase acquired by the environment spins onto the probe spin before readout. 
Here we provide more details of the method presented in Ref.~\cite{Goldstein11} and consider several extension of the work. To this end we assume that the spin environment can be collectively controlled and partially polarized. These spins could thus be considered as an ancillary system. Still, since they cannot be addressed individually nor read out, they cannot be used directly as probes or in sequential adaptive schemes~\cite{Schaffry11,Giovannetti06,Higgins07}. In addition, because their couplings to the probe spin cannot be switched off, they are a cause of decoherence for the probe spin (as we will see in section \ref{Decoherence}) and thus they can be considered as environment. Nevertheless we show that one can make active use of these spins, to increase the sensitivity of a measurement. 

\begin{figure}[t]
\begin{center}
		\includegraphics[scale=0.25]{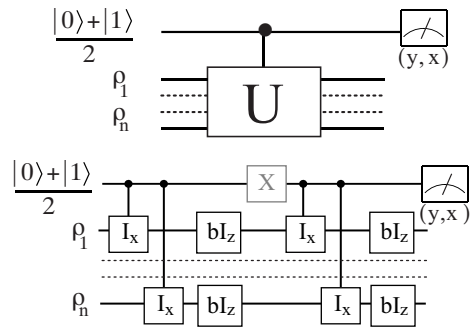}
		\end{center}
	\caption{\label{fig:Circuit} (a) Ideal circuit for EAM with ancillary qubits, based on a DQC1 scheme.  (b) More realistic circuit, where the interaction with the field to be measured is not conditional on the probe spin. The controlled-$I_x$ gates denote C-Not gates, in the ideal model, and are reduced to more general rotations (with different angles for different spins) in the realistic scheme.}
\end{figure}

Ancillary qubits have been considered as a resource for parameter estimation~\cite{Boixo08} in a 
scheme inspired by the deterministic quantum computing with only one pure qubit (DQC1) model~\cite{Knill98}.
  In that scheme,  the probe qubit is initially prepared in a superposition state, then the ancillary system interacts with the external parameter \textit{conditional} on the state of the probe, which is finally readout (see Fig.~\ref{fig:Circuit}.a). When the conditional evolution is given by the operator $U=e^{-ibt\sum_kI_z^k}$ (where $\vec{I}^k$ are the ancilla spin operators) the sensitivity achieves the SQL 
(scaling as 1$/\sqrt{n}$ where $n$ is the number of ancillary qubits) for ancillas in a completely mixed state~\cite{Boixo08} and the Heisenberg limit for pure state (scaling as $1/n$). In that case, it is convenient to read out the y-component of the probe spin, which gives a signal $S=\sin(nbt)$.  
Since the signal is enhanced by a factor of $n$ for small fields $nbt\ll1$ this  yields an Heisenberg-limited sensitivity scaling as $1/n$. Indeed, for pure input states, the circuit creates an entangled state that provides a signal enhancement. 
Below we modify this scheme so that it can be implemented for realistic physical systems.

In general, the ancillas dependence on the external parameter cannot be controlled by the probe spin,  
as it is implicitly assumed above. Thus it is necessary to intersperse the evolution under the interaction with the external field with C-Not gates (Fig.~\ref{fig:Circuit}.b). With this modification we achieve a similar evolution as before. However,  even this simpler scheme cannot be easily implemented and is not compatible with our assumptions of limited control on the environment spins: if the ancillas are spins in the environment, it is not possible to control them individually, thus the C-Not gates cannot be implemented since the required interaction time for the C-Not operations will be different for the different spins. 
The key to using the environment spins -- with the corresponding limited control -- as a resource for parameter estimation is to realize that the scheme works also if the the controlled gates are not ideal $\pi$-rotations. 
The rotations can differ for different spins, as long as the state of the probe spin is flipped (Not gate) before the second set of controlled gates: this ensures that all the environment spins contributes constructively to the final phase, 
as we derive below.  

We note that the spin flip of the probe achieves two other results: first, it makes the evolution insensitive to static noise (as produced for example by a very slowly varying spin bath) since the gate amount to a spin echo for the probe spin.
Secondly, the echo pulse refocuses the entanglement created in the first half of the circuit; this operation cancels undesired terms in the signal that would arise when considering a more realistic scenario where both the external field and the couplings to the probe spin used to create controlled rotations are always present at the same time.

\begin{figure}[t]
\centering
\includegraphics[scale=0.9]{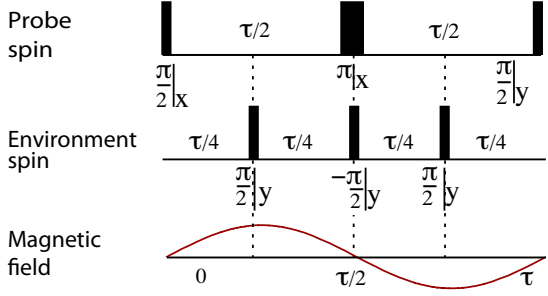}
\caption{EAM pulse sequence: the vertical bars represent microwave pulses on resonance with the probe (top part of the figure) or environment spins (center), performing the labeled rotations. We assume that the field to be measured is an AC field synchronized with the pulse sequence 
as shown in the bottom of the figure.}
\label{fig:Sequence}
\end{figure}

The idealized scheme in Fig.~\ref{fig:Circuit} can be implemented in practice with realistic resources, with the EAM pulse sequence of  Fig.~\ref{fig:Sequence}. We  consider a system comprising a sensor spin ($S=1$) and environment spins
($I^k$), which in a convenient rotating frame on resonance with the $m_s=0\to1$ transition, is described by the Hamiltonian: 
\begin{equation}
\begin{array}{l}
\ham=b(t)\left(\gamma_SS_z+\gamma_I \sum_k I_z^k\right)+\sum\lambda_k S_z I_z^k\\
\quad=\ku\bu \left[b(t)\gamma_I\sum_k I_z^k\right]+\\
\qquad\kd\bd \left[\gamma_S b(t)+\sum_k(\gamma_I b(t)+ \lambda_k) I_z^k\right],
\end{array}
\label{eq:Hamiltonian}
\end{equation}
where $b(t)$ is the external field to be measured, $\gamma_{S,I}$ are the gyromagnetic ratios of the probe and environment spins respectively, 
 $\lambda_{k}$ are the dipole couplings between the sensor and environment spins, and $\ku$ ($\kd$) denotes the $m_s=0$ ($m_s=1$) eigenstate of the $S_z$ operator.

We choose a spin-1 system for its analogy with  Nitrogen-Vacancy centers in diamond~\cite{Jelezko06,Childress06} as they have emerged as good quantum probes of magnetic fields~\cite{Taylor08,Maze08,Balasubramanian08} for their controllability, optical readout and long coherence times. In addition Nitrogen paramagnetic impurities (P1 centers~\cite{Hanson08}) can act as the environment spins, since they can be collectively controlled~\cite{Delange10}. 
The choice of a spin-1 system is in addition important since the presence of an eigenstate with zero eigenvalue effectively allows shutting off the interaction between the probe spin and the environment spins at given times: this flexibility makes the EAM scheme easier to implement. We will lift this restriction and examine more general case in section~\ref{Spins}.

In the sequence in Fig. \ref{fig:Sequence} the probe spin undergoes a spin-echo sequence induced by pulses on resonance with the transition between the states $|0\rangle$ and $|1\rangle$ before being measured. For any given evolution of the environment, the signal can then be calculated from $S(t)=[1+\mathcal{S}(t)]/2$, with~\cite{Witzel06,Rowan65}:
\begin{equation}
\mathcal{S}(t)=\text{Im}\left[\tr{U_0U_1\rho_{\text{env}}U_0^\dag U_1^\dag}\right]
\label{eq:pseudospin}
\end{equation}
Here the propagators $U_i=e^{-i\ham_it}$ are defined as the evolution of the environment spins in the $m_s=i$ manifold, where $\ham_0=b(t)\gamma_I\sum_k I_z^k$ and $\ham_1=\gamma_S b(t)+\sum_k(\gamma_I b(t)+ \lambda_k) I_z^k$ (see Eq.~\ref{eq:Hamiltonian}).
The pulsed evolution of the environment, giving the propagators $U_i$, can be most easily calculated in the \textit{toggling} frame~\cite{Haeberlen76}, the interaction frame defined by the control pulses. In this frame, the Hamiltonian (\ref{eq:Hamiltonian}) becomes piecewise time-dependent, with operators alternating between the z- and x-axis.

The evolution  for the sequence of Fig.~\ref{fig:Sequence} and the resulting signal (Eq.~\ref{eq:pseudospin}) can be calculated exactly in the case of a single ancilla. Here we will present only the result for many ancillas in the limit of \textit{small} field $b$, following the derivation of 
Ref.~\cite{Goldstein11}. We neglect for the moment the coupling of the sensor spin to the magnetic field and 
only keep first order term in the field.   By expanding the exponentials,  the only terms  contributing to the signal are then
\[
\begin{array}{l}
\text{Im}\left[
\text{Tr}\left\{
e^{-i\tau/4\sum_k\lambda_kI^k_x}e^{-i\tau/4\sum_k\lambda_kI^k_z}\rho_{env}e^{i\tau/4\sum_k\lambda_kI^k_z}\times\right.\right.
\\
\left.\left.
e^{i\tau/4\sum_k\lambda_kI^k_x}
({-i}\overline{B}_2\tau\sum I^k_z)
\right\}\right]=- \overline{B}_2\tau\sum_k\cos(\lambda \tau/4)
\end{array}
\]
and
\[
\begin{array}{l}
\text{Im}\left[
\text{Tr}\left\{
e^{-i\tau/4\sum_k\lambda_kI^k_x}e^{-i\tau/4\sum_k\lambda_kI^k_z}(i  \overline{B}_2\tau\sum I^k_z)\rho_{env}\times\right.\right.
\\
\left.\left.
e^{i\tau/4\sum_k\lambda_kI^k_z}
e^{i\tau/4\sum_k\lambda_kI^k_x}
\right\}\right]=\overline{B_{2}}\tau,
\end{array}
\]
 inserted dt in integrals
where $\overline{B_{2}}=-\frac{1}{\tau}\int_{\frac{\tau}{2}}^{\frac{3\tau}{4}}b\left(t\right) dt$.\\
The signal is then given by $S=\half[1-\sin(\Phi)]$, with 
\begin{equation}
\Phi=\gamma_S\overline{B_{1}}\left[1+2P\frac{\gamma_I\overline{B_{2}}}{\gamma_S\overline{B_{1}}}\sum_k \sin\left(\frac{\lambda_{k}\tau}8\right)^2\right],
\label{eq:Phase}
\end{equation}
where $\overline{B_{1}}=\frac{1}{\tau}\left(\int_{0}^{\frac{\tau}{2}}b(t)dt-\int_{\frac{\tau}{2}}^{\tau}b(t)dt\right)$ is the contribution from the direct coupling of the sensor with the field  and we 
have introduced the  polarization $P\leq1$ of the environment spins, so that the initial state of each spin in the environment is $\rho_k=\Id/2+PI_z^k$. 
The factor in the square bracket is the amplification attained as compared to magnetometry performed via a spin echo~\cite{Taylor08}.  We can always get an amplification, as $\sin\left(\frac{\lambda_{k}\tau}8\right)^2$ is non-negative and changing the pulse phases  always ensures that  $\gamma_I\overline{B_2}$ and $\gamma_S\overline{B_1}$ have the same sign.

\begin{figure*}[t]
\centering\includegraphics[scale=0.32]{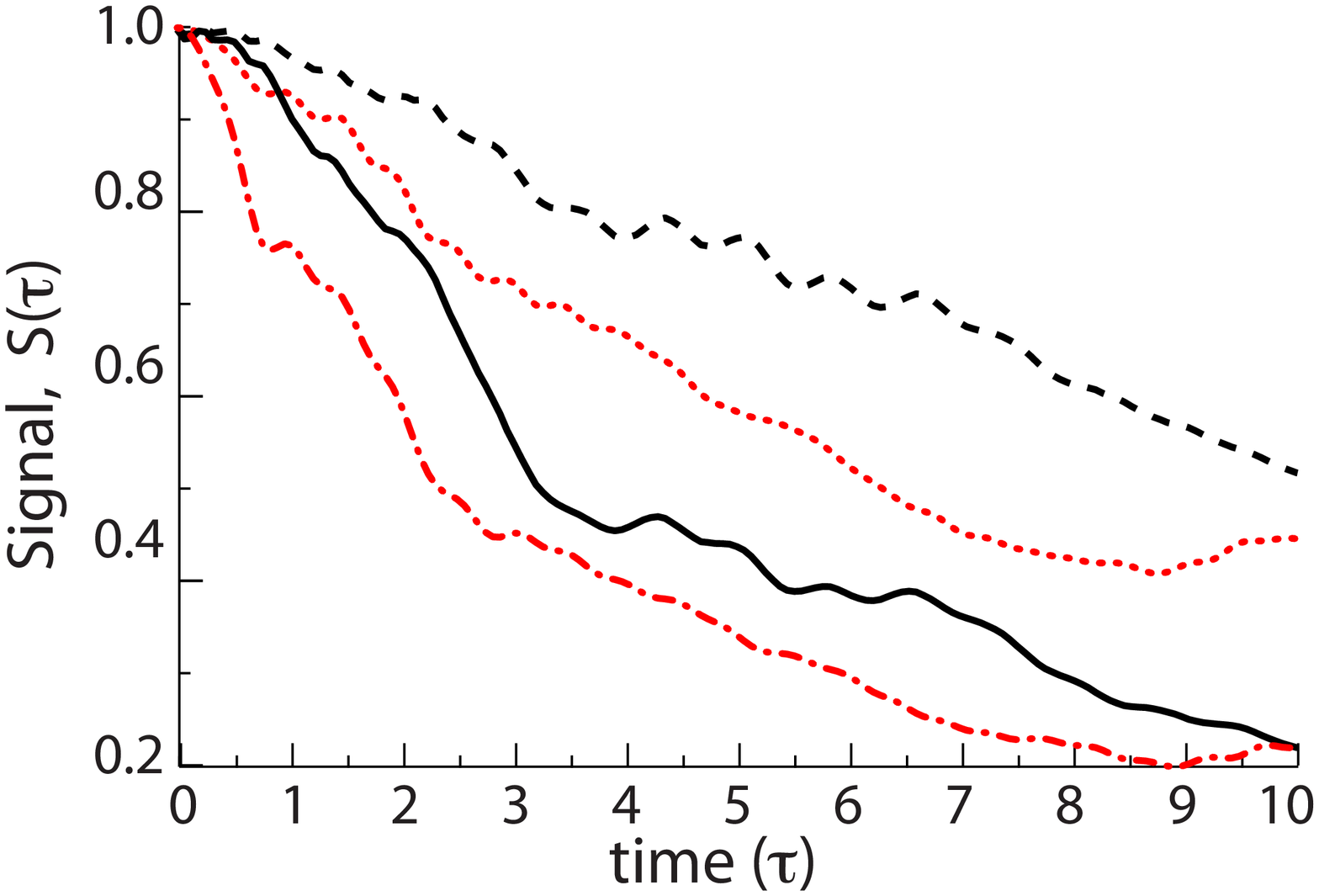}\hspace{42pt}
\includegraphics[scale=0.32]{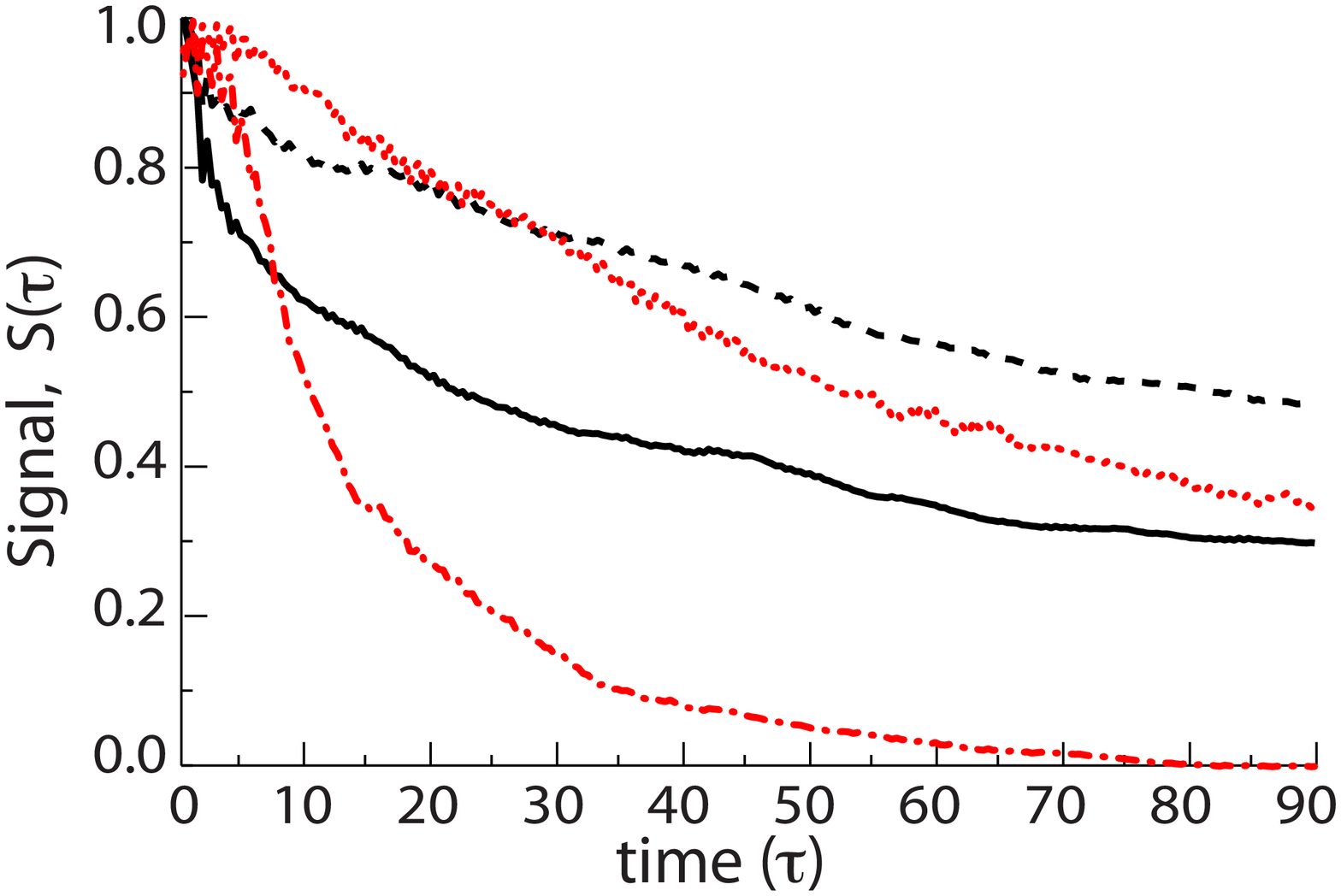}
\caption{
Left: Simulations of  signal decay for  spin echo sequence (Red, dotted and dash-dotted lines) and EAM  sequence (Black, dashed and solid lines).  Right: Simulations of the two sequences with a WAHUHA sequence embedded in each time period (for 1 to 50 cycles). The time was normalized by the largest sensor-environment spin coupling, $\tau\sim[\pi/\lambda_{\rm max}]$.\\
The dotted and dashed lines correspond to a 6\% spin density  and the dash-dotted and solid lines to a density $\approx1/8$. In the first case about 25 environment spins were placed around the probe spin on a diamond lattice, while in the second case, about 50 spins were simulated. The polarization of each environment spin was $P=\half$. 
We took an average of 100 spin distributions to obtain mean decay values and performed each simulation using the disjoint cluster method~\cite{Maze08b}, with clusters of 6 spins.}
\label{fig:Simulation}
\end{figure*}

For values of the couplings such that $\left|\lambda_{k}\tau\right|\gtrsim2\pi$, or \textit{strongly coupled} environment spins, the terms $\sin\left(\frac{\lambda_{k}\tau}8\right)^2$ average to $\half$. \textit{Weakly coupled} environment spins ($\lambda_{k}\leq1$) contribute instead with a factor $\propto\lambda_{k}^{2}$ and we obtain a total phase 
\begin{equation}
\Phi=\gamma_S\overline{B_{1}}\tau\left[1+ P\frac{\gamma_I\overline{B_{2}}}{\gamma_S\overline{B_{1}}}\left(n_{sc}+2\sum \,^{'}(\lambda_k\tau/8)^2\right)\right],
\label{eq:PhaseStrong}
\end{equation}
where $n_{sc}$ is the number of strongly coupled spins and the primed sum runs only over the weakly coupled spins (this last term can generally be neglected compared to the strongly coupled spin contribution).

The sensitivity of the EAM scheme is easily calculated by noting that ideally the only noise contribution is the shot noise of the spin probe. For $\gamma_S=\gamma_I\equiv\gamma$ and assuming an oscillating field  in phase with the echo sequence
$b(t)=b_0\sin(2\pi t/\tau)$,  the sensitivity~\cite{Bollinger96,Wineland94} per unit time  $\eta=\frac{\Delta S}{\|\pdev S{b_0}\|}\sqrt{T}$ is
\begin{equation}
\eta=\frac{\pi}{C\gamma(2+\half P n_{sc})\sqrt{\tau}},
\label{eq:SensitivityIdeal}
\end{equation}
where we introduced the factor $C$~\cite{Taylor08} to include any non-ideality of the measurement procedure (here we assumed $T=N\tau$, with $N$ the number of repetitions of the measurement). The sensitivity scales as $1/n_{sc}$ achieving a Heisenberg-like scaling\footnote{Equation~(\ref{eq:SensitivityIdeal}) is valid only for $P\neq0$. We will analyze the case $P=0$ in Section~\ref{Polarization}.}.

We note that even in this ideal case, there are two factors that reduce the sensitivity:   
a limited polarization of the environment spins and the reduction of the  time during which the interaction with the external field is effective (because of the scheme proposed, a phase is acquired which is proportional to only $1/4^{th}$ of the total sequence time).

The EAM scheme thus demonstrates that it is possible to
attain nearly Heisenberg limited sensitivity for metrology with a
new class of entangled states (other than squeezed or GHZ states) that as we
will see in the following are more robust to decoherence. 
Furthermore, these states can be created with limited control resources, thus opening the possibility of using spins in the environment as a resource for metrology.

\section{Decoherence}\label{Decoherence}
The results in the previous section did not take into account the effects of decoherence caused both by the environment spins used as an ancillary system and by any other residual bath. In this section we will take these effects into account and   show that even in the non-ideal case the EAM sequence can provide a sensitivity enhancement with respect to other control scenarios (such as a spin-echo) that only aim at refocusing the interaction of the probe spin with the environment spins.

\subsection{Decoherence induced by the environment spins couplings}
The  interactions among environment spins  hamper the EAM scheme in two ways.
First, flip-flops of environment spins lead to a loss of coherence of the probe spin. 
This effect is the same that is observed during a spin echo, and we will show that the resulting coherence time $T_2$ is on the same order for the two sequences. Second, the interactions will also cause the environment spins to lose their internal phase coherence, resulting in a smaller accumulated phase $\Phi$. Still, this effect happens on a time scale $\tau_I$ given by the environment spin correlation time, which is usually longer than the probe coherence time, $\tau_I\geq T_2$. Thus the sensitivity is ultimately limited by $T_2$, as in the spin-echo case.

Consider the system evolution as given by Eq.~\ref{eq:pseudospin} (for simplicity in the absence of the magnetic field $b$). Now the propagators are given by the Hamiltonian
\begin{equation}
\begin{array}{l}
H= b\left(t\right)\left(\gamma_SS_{z}+\gamma_I \sum I_z^k\right)+\sum\lambda_kS_zI_z^k\\
\qquad+\sum\kappa_{jk}\left(3I_z^jI_z^k-\vec{I^j}\cdot\vec{I^k}\right)
\end{array}
\label{eq:FullHamiltonian}
\end{equation}
where $\kappa_{ij}$ are the intra-bath couplings given by the magnetic dipole interaction among spins.
Because of the presence of the couplings,  the evolution in the two halves of the sequence is no longer the same, thus the interaction  between the probe spin and the environment spins can no longer be perfectly refocused. This effect, usually called spectral diffusion, is observed as well in spin echo experiments and lead to the coherence time $T_2$. The addition of a modulation of the environment spins is not expected to change substantially the coherence time, as hinted by 
the short time evolution expansion presented in Ref.~\cite{Goldstein11}. An exception is for a perfectly polarized bath: in that case, flip-flops are quenched in the spin-echo, but they are still allowed in the EAM scheme  
since they are enabled by the rotation of the spins during the protocol; the effect of flip-flop quenching is however noticeable only for very high polarization of the bath~\cite{Fischer09,Takahashi08}. 

From this argument we expect that one can have a similar interrogation time $\tau$ in Eq. (\ref{eq:SensitivityIdeal})  for the EAM scheme considered here as for  a simple spin echo sequence. Unlike for different entangled states~\cite{Huelga97}, 
the enhancement from entanglement is therefore not counterbalanced by a decrease in the interrogation time $\tau$, and the EAM scheme does allow for a significant improvement of the sensitivity. 

We further verify this claim by simulations. We used the disjoint cluster approximation~\cite{Maze08b} to simulate the sequence in Fig.~\ref{fig:Sequence} for a system comprising the probe spin surrounded by an environment of 25-50 spins randomly positioned in a  
cube with sides of unit length. 
By averaging over many spatial distributions of the environment spins, the simulation converges quickly even for small cluster sizes and it gives information about the average coherence time~\cite{Yang08,Witzel10}.

The system we consider is inspired by a NV center in a nano-crystal of diamond in the presence of P1 Nitrogen impurities~\cite{Goldstein11}, but the results are more generally valid. For comparison, we also simulated the evolution under a spin echo sequence. From the results in Fig.~\ref{fig:Simulation} we see that the coherence time is not qualitatively different for the two sequences.
The figure shows in addition that the coherence time depends on the density of the environment spins, a fact that will be important in evaluating the sensitivity achievable with the EAM scheme.

The second effect of the intra-bath couplings is to make the environment spin themselves loose their coherence, in a time on the order of their correlation time $\tau_I$, which is given by the rate of spin flip-flop driven by the dipolar interaction. If the environment spins are no longer in a coherent state, the phase they acquire does not add up constructively, resulting in a smaller phase $\Phi$. 
Still, this effect is comparable to the previous one, since the correlation time is at  least on the same order of $T_2$. 

In addition to the environment spins that are used as ancillary sensors, the system could be in contact with an additional spin bath. For example, in the case of the NV center in diamond this bath is given by the \carb nuclear spins. The effects of this quasi-static bath are refocused by the $\pi$ pulse on the NV center and by the two $\pi/2$ pulses on the environment spins, which amount to a so-called ``Hahn echo''~\cite{Hahn50} sequence. Any residual decay is again comparable to what is observed in a simple spin echo for the probe spin.

\subsection{Dynamical Decoupling}\label{sec:decoupling}
 An increase in the effective correlation time of the environment spins would be beneficial in two ways, by both increasing the coherence time of the probe spin, through its influence on the sensor spin $T_2$-time, and directly by improving the environment spin coherence. Dynamical decoupling schemes could achieve this goal.
\begin{figure}[htb]
\centering\includegraphics[scale=0.85]{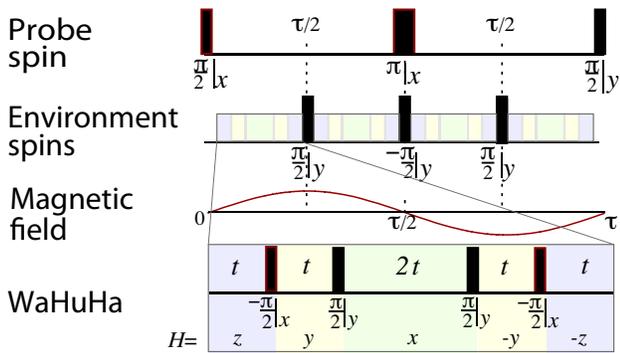}
\caption{ (color online) Embedding of a WAHUHA sequence in the EAM sequence. The WAHUHA is shown at the bottom, together with the ``direction'' of the Hamiltonian in the toggling frame. 
}
\label{fig:WHH}
\end{figure}
The dipolar Hamiltonian can be refocused using homonuclear decoupling sequences such as the WAHUHA sequence~\cite{Waugh68}. The pulse modulation gives a time-dependent Hamiltonian 
for the spin-spin interaction that averages to zero over a cycle time $t_c$. If the modulation is fast compared to the couplings, the effective Hamiltonian over the cycle is well approximated by its average. A simple symmetrization of the pulse sequence~\cite{Burum79}  
can further cancel out the first order correction, leaving errors that are only quadratic in the product $\kappa t_c$ (and do not depend on the total evolution time, that could be given by many cycles)~\cite{Haeberlen76}.

Fig.~\ref{fig:WHH} shows how to incorporate a WAHUHA sequence within the EAM  sequence. 
We modified the phases of the pulses with respect to the original sequence in order to obtain  an effective coupling between the probe and environment spins  $\propto\frac{1}{\sqrt{3}}S_{z}\sum\lambda_{k}I_{z(x)}^{k}$ in the odd(even) time intervals. 
These phase changes do not affect the average of the dipolar Hamiltonian and hence the performance of the WAHUHA sequence. 
Unfortunately the modulation does not only averages out the dipolar Hamiltonian, but it also reduces the linear terms, by a factor $1/\sqrt{3}$. In many cases, the increase in coherence time more than compensate for this weighting factor. In Fig.~\ref{fig:Simulation} we simulated via the disjoint cluster method the coherence of the EAM and spin-echo sequences, while applying the WAHUHA sequence in between the pulses. 
Comparing the results obtained in the absence of dynamical decoupling, we see that the sequence is very effective in increasing the coherence time.

A different strategy for directly increasing the probe spin $T_2$ is to use more than one $\pi$-pulse during the total sequence time~\cite{Taylor08}. This technique is inspired  
by concatenated dynamical decoupling schemes and in particular by the CPMG sequence~\cite{Carr54,Meiboom58}.
More generally, these examples indicate that the EAM scheme can be combined with various forms of decoupling.

\section{Sensitivity}\label{Sensitivity}
In the previous section we saw that the coherence time (and hence the time during which the phase can be acquired) depends on the density of the environment spins. For the EAM sequence, the signal too depends on the environment spin density, since it determines how many environment spins are close enough to the probe spin to be considered ``strongly coupled''. 
Thus the optimal sensitivity arises from a compromise between  the environment spin density and the interrogation time. Including the probe decoherence due to the environment spins, as well as other bath contributions, yielding a coherence time $T_2^B$, the sensitivity of Eq.~\ref{eq:SensitivityIdeal} becomes:
\begin{equation}
\eta=\frac{\pi e^{(\tau/T_2)^3}e^{(\tau/T_2^B)^3}}{C\gamma\sqrt{\tau}(2+\half P n_{sc})}
\label{eq:SensitivityReal}
\end{equation}
The functional form we assumed for the decay is inspired by the measured behavior of NV centers in diamond~\cite{Childress06,Delange10} and usually arises from a Lorentzian spectrum of the bath. 
\begin{figure}[b]
\centering
\includegraphics[scale=0.9]{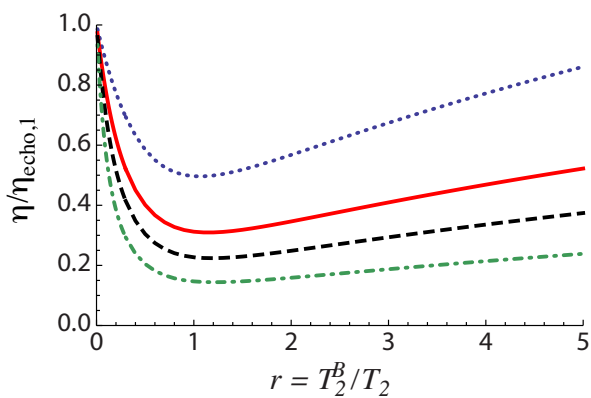}
\caption{ (Color online) Sensitivity  of the EAM scheme normalized by the sensitivity of a spin-echo scheme  in the absence of any environment spin, $\eta/\eta_{\text{echo},1}$. The curves corresponds to  quality factors $Q = 10$ (dotted), 20 (solid), 30 (dashed) and 50 (dash-dotted). 
The ratio improves until $r=T_2^B/T_2=1$, where the decoherence induced by the added environment spins overtakes the background decoherence.}
\label{fig:Sensitivity}
\end{figure}

The couplings between the probe and environment spins scales as $\lambda_k\sim\gamma^2/r_k^3$ (assuming dipolar interaction and setting $\gamma_I=\gamma_S=\gamma$ for simplicity), with $r_k$ the distance to the probe spin. Then, for a fixed duration $\tau$ of the EAM sequence, the number of ``strongly coupled'' spins $n_{sc}$ scales as $n_{sc}(\tau)\sim\gamma^2\rho \tau$, where $\rho$ is the density of the environment spins. 
The probe coherence time also scales with the density as $T_2\propto1/\rho$. 

The sensitivity is then a function of two parameters: how many polarized spins are strongly coupled in the coherence time $T_2$ and how much the coherence time is reduced with respect to the background bath coherence time by introducing the ancillary environment spins. We define 
a quantity $Q=P \rho  \gamma^2 T_2$, which describes the ``quality'' of the environment spins. A second quantity describing the reduction in coherence time due to the ancillary environment spins is  given by the ratio $r=T_2^B/T_2$. 

The EAM sensitivity then depends only on these two parameters and the bath coherence time,  
such that Eq.~(\ref{eq:SensitivityReal}) becomes
\begin{equation}
\eta=\frac{\pi e^{(1+r^3)(\tau/T_2^B)^3}}{C\gamma\sqrt{\tau}(2+\frac\tau{2T_2^B} r Q )}
\label{eq:SensitivityParams}
\end{equation}
We can further optimize the sensitivity with respect to the interrogation time $\tau$ and compare it to the 
case where the field is measured by a probe spin (via a spin-echo sequence) in the presence of the background spin bath only (that is, no environment ancillary spins). In this case, the sensitivity is given by~\cite{Taylor08}
\begin{equation}
\eta_{\text{echo,1}}=\frac{\pi e^{(\tau/T_2^B)^3}}{2C\gamma\sqrt{\tau}}
\label{eq:etaAC}
\end{equation}
\begin{figure}[b]
\centering
\includegraphics[scale=0.9]{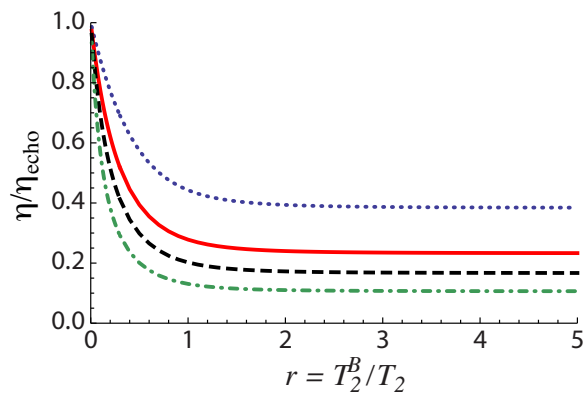}
\caption{(Color online) Sensitivity ratio $\eta/\eta_{\text{echo}}$ between EAM and spin echo schemes as a function of
$r=T_2^B/T_2$  for quality factor $Q = 10$ (dotted), 20 (solid), 30 (dashed), 50 (dash-dotted). When the decoherence induced by the added environment spins dominates (large $r$) the sensitivity ratio is only determined by the environment quality $Q$.}
\label{fig:SensitivityRatio}
\end{figure}

As show in Fig.~\ref{fig:Sensitivity}, the EAM sensitivity  
as given by Eq. (\ref{eq:SensitivityParams}) improves up to $r=1$, where the decoherence  due to the environment spins  becomes more important than the background bath. The improvement depends on the ``quality'' $Q$, since for higher $Q$ there are more strongly coupled spins at a given $T_2$ time. It is then clear that there is an optimum number of environment spins  that one would want to introduce in the system to obtain the optimal sensitivity. Alternatively, the quality $Q$ can be improved by increasing $T_2$ using the dynamical decoupling methods we introduced in section~\ref{sec:decoupling}.

If the number of environment spins is instead fixed, we are interested in comparing the EAM  and  spin-echo scheme for a given system (e.g. a given nanocrystal of diamond).
In Fig.~\ref{fig:SensitivityRatio} we plot  the ratio of the EAM sensitivity to the spin-echo sensitivity :
\begin{equation}
\eta_{\text{echo}}=\frac{\pi e^{(1+r^3)(\tau/T_2^B)^3}}{2C\gamma\sqrt{\tau}}
\label{eq:sensitivityechoenv}
\end{equation}
(in the figure this expression is optimized with respect to the interrogation time $\tau$).
 In the high $r$  limit the sensitivity  ratio depends only on the $Q$ factor, as $\eta/\eta_{\text{echo}}\approx\frac{\sqrt[6]{e^2/3}}{ \left(1+2^{-7/3} Q\right)}$.

To estimate the potential sensitivity improvement of the EAM method we express $Q$ in terms of measurable quantities. 
Specifically, we can write the environment quality as $Q=P\rho\gamma^2T_2\approx P\sqrt{\sum\lambda_k^2} T_2$, where we used the fact that $\gamma^2\rho\tau=n_{sc}(\tau)\lesssim\tau\sqrt{\sum\lambda_k^2} $ . The average distribution of couplings $M_2=\sqrt{\sum\lambda_k^2}$  is related to the second moment of the probe spin, which gives its dephasing time $M_2=1/T_2^*$. Then the sensitivity improvement is given by the ratio $T_2/T_2^*$, which can be quite large in many system.

\section{Polarization and sensitivity}\label{Polarization}
The sensitivity discussed in the previous section depends on the polarization of the environment spins. In this section we first propose  methods for creating this polarization, under the assumption that the probe spin can be polarized at will. We then generalize the EAM scheme to the case where no polarization is available. This generalization will furthermore prove useful in the case where the field to be measured is affected by a random phase.

\subsection{Polarizing the environment spins}
In an environment assisted magnetometer working at room temperature, the environment spins will be in a thermal state, close to the maximally mixed state. Polarization need then to be created by relying on the probe spin and the Hamiltonian (Eq.~\ref{eq:FullHamiltonian}) that is required for the measurement scheme.  
To do this we assume that the probe spin can be repetitively polarized: this is the case e.g. for an NV center that can be polarized optically. Polarization could then be  transferred to the spins in the environment by a swapping Hamiltonian such as $\ham_{SW}\sim (S_xI_x+S_yI_y)$. Although this operator is contained in  the dipole-dipole Hamiltonian, it is usually quenched in the rotating frame, if the energies of the two spin species are different. For example, in the case of NV and P1 spins, the zero-field splitting of the NV creates an energy mismatch. 

The swapping Hamiltonian can be reintroduced by inducing a Hartman-Hahn matching of the energies in the rotating frame under a continuous microwave irradiation~\cite{Hartman62,Weis06}. By adjusting the Rabi frequency and the offset, the two spin species are brought into resonance and spin flip-flops (allowing polarization transfer) are now allowed, 
leading to a buildup of polarization. 
The environment spins can then be polarized efficiently by alternating periods during which the probe spin is polarized and periods during which polarization exchange is driven by the microwave irradiation.

The buildup of polarization can happen either via direct interaction between the probe spin and a spin in the environment, or indirectly via spin-diffusion~\cite{Khutsishvili66,Ramanathan08}. Since we are interested only in polarizing strongly coupled spins, the first process is dominant. 
Then, we can estimate the polarization time by the number of spins we want to polarize divided by their average coupling strength, $T_{pol}\sim n_{sc}(T)/\ave{\lambda}\approx n_{sc}(T)T_2^*$ (where we used $1/T_2^*=\sqrt{\sum\lambda_k^2}$ to estimate the average coupling strength, an upper bound for $T_{pol}$ would be more generally   $T_{pol} \leq n_{sc}(T) T/\pi$).

A different strategy to initialize the spin environment is  measurement-based polarization~\cite{Togan11} with either  feedback or adaptive schemes. Precise measurement of the local magnetic field created by the spin environment at the sensor spin location effectively determines the environment spin state, with an increasing knowledge of the magnetic field shift corresponding to a reduced spin-state distribution and hence higher polarization.    

The polarization time will reduce the achievable sensitivity per root Hz, $\eta$, since it increase the preparation time 
such that fewer measurement can be performed during a certain time interval. The exact sensitivity degradation  will depend on many factors, e.g. the depolarization ($T_1$) time of the environment spins, which determines how often the preparation step needs to be repeated.

\subsection{EAM with no polarization and phase error}
In the discussion so far we assumed that 
the external field to be measured was either static or oscillating in phase with the control sequence.
For $b(t)=b\cos(2\pi t/\tau)$,  we obtained the signal $S=\half(1-\sin\Phi)$, where the phase is given by Eq.~(\ref{eq:Phase}) for the EAM scheme or by $\Phi=2\gamma b \tau/\pi$ for the spin-echo scheme. However, if the field has a random phase  (or cannot be synchronized perfectly with the pulse sequence) the signal averaged over many runs goes to zero, as $\ave{S}=\half(1-\ave{\sin\Phi})=0$ if $\ave{\Phi}=0$.   
 Furthermore, even higher momenta of the signal, $\ave{S^n}$ are zero; thus it is not possible to infer information about the stochastic field by this method.
 
 A possible solution is to change the phase of the final pulse~\cite{Meriles10} (or equivalently, to introduce an additional, known phase accumulation during the free evolution). 
 Then the signal becomes
 \begin{equation}
S=\half[1+\ave{\cos(\Phi+\theta)}]
\label{eq:SignalNoPol}
\end{equation}
where $\theta$ is the phase difference between the initial and final pulse, $\Phi$ is the phase due to the field to be measured and we neglect any decay for simplicity. 
Since $\ave{\cos(\Phi+\theta)}=\ave{\cos\Phi}\cos\theta$, the maximum signal is obtained for $\theta=0$, or by setting the phase of the initial and final pulse to be equal.

The phase $\Phi$ acquired in the modified EAM scheme (Fig.~\ref{fig:Sequence} with the last pulse along x) is different than 
that obtained in Eq.~(\ref{eq:Phase}). In the limit of small fields, we obtain the signal
\begin{equation}\label{eq:SignalEAMNoPol}\begin{array}{l}
S_x=1-\half(\frac{b t}{2\pi})^2\left(2+ \sum_k[1+\cos\left(\frac{\lambda_k t}4\right)^2] \sin \left(\frac{\lambda_k t}4\right)^2\right)\\
\quad \approx 1-\half\left(\frac{b t}{2\pi}\right)^2\left[2+ \frac34 n_{sc}\right],
\end{array}
\end{equation} 
where again we only summed over the ``strongly coupled'' environment spins. 
We note that the signal does not depend on the polarization of the environment spins (at least to first order in the polarization and to second order in the field $b$). Thus, even in the absence of any polarization it is possible to measure the external field (although not the sign of it).

We compare the achievable sensitivity of the EAM  and  spin-echo method in the case where no polarization is present and the control sequence describe above is used. Optimizing the sensitivity with respect to the interrogation time $\tau$, we obtain the sensitivity for the spin echo sequence
\begin{equation}
\eta_{\text{echo},x}= \frac{\pi}{C\gamma\sqrt{\tau}}
\label{eq:sensitivityechox}
\end{equation}
while for the EAM sequence we have
\begin{equation}
\eta_{\text{EAM},x}\approx \frac{\pi }{C\gamma\sqrt{\tau}\sqrt{1+\frac32 n_{sc}}}
\label{eq:sensitivityeamx}
\end{equation}
In the case of zero polarization (or of a signal with a random phase) it is no longer possible to obtain a quantum enhancement and have a scaling proportional to $1/n_{sc}$ by exploiting the spins in the bath. Indeed if there is no polarization, no entanglement is created in the system, and no quantum enhancement of the sensitivity is expected\footnote{Note that  our assumptions of no direct access to individual ancillary spins preclude using  adaptive schemes,  which have been shown in other conditions to achieve quantum enhancement of the sensitivity even without entanglement~\cite{Higgins07}.}. 
Nevertheless, for favorable conditions of the spin environment (high quality $Q$ and low ratio $r$) it might still be beneficial to use the EAM scheme instead of a simple spin-echo magnetometry 
because it allows for an improvement  $\sim\sqrt{n_{sc}}$ by exploiting the unpolarized spins. 

\section{Extension to other spin probes}\label{Spins}
In the previous sections we presented a scheme that relied on the fact that for one of the eigenstates of the probe spin ($\ku$) the couplings to the environment spins was zero.  
It is possible to extend the EAM scheme to the case where the probe is a spin-1/2,  but only if 
the environment-probe spin couplings are all of the same sign. 
Such a situation could for example be realized by considering a single quantum dot in the nuclear spin environment.
The interaction between the central spin and the environment spins  
in this system is given by the contact interaction, whose strength depends mainly on the electronic spin wavefunction density and does not present the strong angular dependence of the dipolar interaction.

To apply the EAM scheme with a spin-$1/2$ probe, we rotate the environment spins to be aligned along the $y$ axis before applying the sequence shown in Fig. (\ref{fig:Sequence}). 
For small fields, the additional phase acquired thanks to the environment spins is given by
\begin{equation}
\Phi_{1/2}\propto b \tau P\sum_k\sin\left(\frac{\lambda_k\tau}4\right)^3
\label{eq:Phase12}
\end{equation}
If all the couplings $\lambda_k$ are positives, the environment spins contributions add constructively and it is always possible to find a time $\tau$ s.t. there are $n_{sc}$ strongly coupled spins for which $0\leq\lambda_k \tau\leq4\pi$, so that $\sum_k\sin\left(\frac{\lambda_k\tau}4\right)^3\approx \frac{4}{3\pi}n_{sc}$.

More generally, it is also possible to use probes with higher spins, selecting two of their eigenstates as the levels of interest, by driving transitions on resonance with their energy difference. If the two eigenstates $\ket{a}$ and $\ket{b}$ are such their eigenvalues have different absolute values, $|m_a|\neq|m_b|$ then we can apply the EAM sequence for any value of the couplings, as the phase enhancement will be $\Phi\propto \sum_k\sin\left(\frac{|m_a|-|m_b|}8\lambda_k \tau\right)^2$. Otherwise, one might use the modified scheme just presented in this section, if all the  
coupling constants are positive.

As shown in this section, the scheme we introduced is quite flexible and can be applied to many different physical systems, beyond the one we focused on in this paper. Besides spin systems, the same ideas could for example find an implementation based on trapped ions~\cite{Goldstein11}.\vspace{-6pt}
\section{Discussion and conclusion}
In conclusion,  we  
 analyzed the EAM scheme introduced in Ref.~\cite{Goldstein11},  which aims at enhancing the sensitivity of a single solid-state spin magnetic field sensor,  by exploiting the possibility to  coherently control part of its spin environment. 
The environment spins act as sensitive probes of the external magnetic field, and their acquired phase is read out via the interaction with the sensor spin.
Since the measurement scheme maintains roughly the same coherence times of spin-echo-based magnetometry and the noise is still the shot-noise of a single qubit,  we achieve a quasi-Heisenberg limited sensitivity enhancement. 
We analyzed in detail the sources of decoherence and confirmed with numerical simulations that the sensor coherence time under the EAM scheme is comparable to the $T_2$ time under spin-echo, since the leading cause for decoherence has the same origin in the two cases. We further showed that dynamical decoupling schemes aimed at increasing the correlation time of the spin environment, by reducing the effects of intra-bath couplings, can be embedded in the measurement scheme and leads to longer coherence times and enhanced sensitivity. 
We extended the EAM scheme to the case where the environment spins are in a highly-mixed (zero-polarization) state, by appropriately modifying the detection sequence. This modified scheme achieve a classical scaling of the sensitivity, but can still be beneficial whenever the polarization methods we outlined cannot be applied or the AC field to be measured has a random phase. 
Our analysis finds that the sensitivity  is determined by the ``quality'' of the environment, a parameter that takes into account a compromise between the number of strongly coupled environment spins with the reduced coherence time they entail. This result can be used to define the specifications of  engineered systems with controlled densities of spin impurities for optimal sensitivity.  

 \textbf{Acknowledgments}. This work was supported by the NSF,  NIST, ARO (MURI - QuISM), DARPA QuASAR, the Packard Foundation and the Danish National Research Foundation.

\bibliographystyle{apsrev4P}

\begin{thebibliography}{10}%
\makeatletter
\providecommand \@ifxundefined [1]{%
 \ifx #1\undefined \expandafter \@firstoftwo
 \else \expandafter \@secondoftwo
\fi
}%
\providecommand \@ifnum [1]{%
 \ifnum #1\expandafter \@firstoftwo
 \else \expandafter \@secondoftwo
\fi
}%
\providecommand \enquote [1]{``#1''}%
\providecommand \bibnamefont  [1]{#1}%
\providecommand \bibfnamefont [1]{#1}%
\providecommand \citenamefont [1]{#1}%
\providecommand\href[0]{\@sanitize\@href}%
\providecommand\@href[1]{\endgroup\@@startlink{#1}\endgroup\@@href}%
\providecommand\@@href[1]{#1\@@endlink}%
\providecommand \@sanitize [0]{\begingroup\catcode`\&12\catcode`\#12\relax}%
\@ifxundefined \pdfoutput {\@firstoftwo}{%
 \@ifnum{\z@=\pdfoutput}{\@firstoftwo}{\@secondoftwo}%
}{%
 \providecommand\@@startlink[1]{\leavevmode}%
 \providecommand\@@endlink[0]{}%
}{%
 \providecommand\@@startlink[1]{%
  \leavevmode
  \pdfstartlink
   attr{/Border[0 0 1 ]/H/I/C[0 1 1]}%
   user{/Subtype/Link/A<</Type/Action/S/URI/URI(#1)>>}%
  \relax
 }%
 \providecommand\@@endlink[0]{\pdfendlink}%
}%
\providecommand \url  [0]{\begingroup\@sanitize \@url }%
\providecommand \@url [1]{\endgroup\@href {#1}{\urlprefix}}%
\providecommand \urlprefix [0]{URL }%
\providecommand \Eprint[0]{\href }%
\@ifxundefined \urlstyle {%
  \providecommand \doi [1]{doi:\discretionary{}{}{}#1}%
}{%
  \providecommand \doi [0]{doi:\discretionary{}{}{}\begingroup
  \urlstyle{rm}\Url }%
}%
\providecommand \doibase [0]{http://dx.doi.org/}%
\providecommand \Doi[1]{\href{\doibase#1}}%
\providecommand \bibAnnote [3]{%
  \BibitemShut{#1}%
  \begin{quotation}\noindent
    \textsc{Key:}\ #2\\\textsc{Annotation:}\ #3%
  \end{quotation}%
}%
\providecommand \bibAnnoteFile [2]{%
  \IfFileExists{#2}{\bibAnnote {#1} {#2} {\input{#2}}}{}%
}%
\providecommand \typeout [0]{\immediate \write \m@ne }%
\providecommand \selectlanguage [0]{\@gobble}%
\providecommand \bibinfo [0]{\@secondoftwo}%
\providecommand \bibfield [0]{\@secondoftwo}%
\providecommand \translation [1]{[#1]}%
\providecommand \BibitemOpen[0]{}%
\providecommand \bibitemStop [0]{}%
\providecommand \bibitemNoStop [0]{.\EOS\space}%
\providecommand \EOS [0]{\spacefactor3000\relax}%
\providecommand \BibitemShut [1]{\csname bibitem#1\endcsname}%
\bibitem{Giovannetti11}%
  \BibitemOpen
  \bibfield{author}{%
  \bibinfo {author} {\bibfnamefont{V.}~\bibnamefont{Giovannetti}}, \bibinfo
  {author} {\bibfnamefont{S.}~\bibnamefont{Lloyd}},\ and\ \bibinfo {author}
  {\bibfnamefont{L.}~\bibnamefont{Maccone}},\ }%
  \bibfield{journal}{%
  \Doi{10.1038/nphoton.2011.35}{\bibinfo {journal} {Nat. Photon}}\ }%
  \textbf{\bibinfo {volume} {5}},\ \bibinfo {pages} {222} (\bibinfo {year}
  {2011})%
  \bibAnnoteFile{NoStop}{Giovannetti11}%
\bibitem{Goldstein11}%
  \BibitemOpen
  \bibfield{author}{%
  \bibinfo {author} {\bibfnamefont{G.}~\bibnamefont{Goldstein}}, \bibinfo
  {author} {\bibfnamefont{P.}~\bibnamefont{Cappellaro}}, \bibinfo {author}
  {\bibfnamefont{J.~R.}\ \bibnamefont{{Maze}}}, \bibinfo {author}
  {\bibfnamefont{J.~S.}\ \bibnamefont{Hodges}}, \bibinfo {author}
  {\bibfnamefont{L.}~\bibnamefont{Jiang}}, \bibnamefont{et~al.},\ }%
  \bibfield{journal}{%
  \Doi{10.1103/PhysRevLett.106.140502}{\bibinfo {journal} {Phys. Rev. Lett.}}\
  }%
  \textbf{\bibinfo {volume} {106}},\ \bibinfo {pages} {140502} (\bibinfo {year}
  {2011})%
  \bibAnnoteFile{NoStop}{Goldstein11}%
\bibitem{Taylor08}%
  \BibitemOpen
  \bibfield{author}{%
  \bibinfo {author} {\bibfnamefont{J.~M.}\ \bibnamefont{Taylor}}, \bibinfo
  {author} {\bibfnamefont{P.}~\bibnamefont{Cappellaro}}, \bibinfo {author}
  {\bibfnamefont{L.}~\bibnamefont{Childress}}, \bibinfo {author}
  {\bibfnamefont{L.}~\bibnamefont{Jiang}}, \bibinfo {author}
  {\bibfnamefont{D.}~\bibnamefont{Budker}}, \bibnamefont{et~al.},\ }%
  \bibfield{journal}{%
  \Doi{10.1038/nphys1075}{\bibinfo {journal} {Nature Phys.}}\ }%
  \textbf{\bibinfo {volume} {4}},\ \bibinfo {pages} {810} (\bibinfo {year}
  {2008})%
  \bibAnnoteFile{NoStop}{Taylor08}%
\bibitem{Schaffry11}%
  \BibitemOpen
  \bibfield{author}{%
  \bibinfo {author} {\bibfnamefont{M.}~\bibnamefont{Schaffry}}, \bibinfo
  {author} {\bibfnamefont{E.~M.}\ \bibnamefont{Gauger}}, \bibinfo {author}
  {\bibfnamefont{J.~J.~L.}\ \bibnamefont{Morton}},\ and\ \bibinfo {author}
  {\bibfnamefont{S.~C.}\ \bibnamefont{Benjamin}},\ }%
  \bibfield{journal}{%
  \Doi{10.1103/PhysRevLett.107.207210}{\bibinfo {journal} {Phys. Rev. Lett.}}\
  }%
  \textbf{\bibinfo {volume} {107}},\ \bibinfo {pages} {207210} (\bibinfo {year}
  {2011})%
  \bibAnnoteFile{NoStop}{Schaffry11}%
\bibitem{Giovannetti06}%
  \BibitemOpen
  \bibfield{author}{%
  \bibinfo {author} {\bibfnamefont{V.}~\bibnamefont{Giovannetti}}, \bibinfo
  {author} {\bibfnamefont{S.}~\bibnamefont{Lloyd}},\ and\ \bibinfo {author}
  {\bibfnamefont{L.}~\bibnamefont{Maccone}},\ }%
  \bibfield{journal}{%
  \Doi{10.1103/PhysRevLett.96.010401}{\bibinfo {journal} {Phys. Rev. Lett.}}\
  }%
  \textbf{\bibinfo {volume} {96}},\ \bibinfo {eid} {010401} (\bibinfo {year}
  {2006})%
  \bibAnnoteFile{NoStop}{Giovannetti06}%
\bibitem{Higgins07}%
  \BibitemOpen
  \bibfield{author}{%
  \bibinfo {author} {\bibfnamefont{B.~L.}\ \bibnamefont{Higgins}}, \bibinfo
  {author} {\bibfnamefont{D.~W.}\ \bibnamefont{Berry}}, \bibinfo {author}
  {\bibfnamefont{S.~D.}\ \bibnamefont{Bartlett}}, \bibinfo {author}
  {\bibfnamefont{H.~M.}\ \bibnamefont{Wiseman}},\ and\ \bibinfo {author}
  {\bibfnamefont{G.~J.}\ \bibnamefont{Pryde}},\ }%
  \bibfield{journal}{%
  \Doi{10.1038/nature06257}{\bibinfo {journal} {Nature}}\ }%
  \textbf{\bibinfo {volume} {450}},\ \bibinfo {pages} {393} (\bibinfo {year}
  {2007})%
  \bibAnnoteFile{NoStop}{Higgins07}%
\bibitem{Boixo08}%
  \BibitemOpen
  \bibfield{author}{%
  \bibinfo {author} {\bibfnamefont{S.}~\bibnamefont{Boixo}}\ and\ \bibinfo
  {author} {\bibfnamefont{R.~D.}\ \bibnamefont{Somma}},\ }%
  \bibfield{journal}{%
  \Doi{10.1103/PhysRevA.77.052320}{\bibinfo {journal} {Phys. Rev. A}}\ }%
  \textbf{\bibinfo {volume} {77}},\ \bibinfo {eid} {052320} (\bibinfo {year}
  {2008})%
  \bibAnnoteFile{NoStop}{Boixo08}%
\bibitem{Knill98}%
  \BibitemOpen
  \bibfield{author}{%
  \bibinfo {author} {\bibfnamefont{E.}~\bibnamefont{Knill}}\ and\ \bibinfo
  {author} {\bibfnamefont{R.}~\bibnamefont{Laflamme}},\ }%
  \bibfield{journal}{%
  \Doi{10.1103/PhysRevLett.81.5672}{\bibinfo {journal} {Phys. Rev. Lett.}}\ }%
  \textbf{\bibinfo {volume} {81}},\ \bibinfo {pages} {5672} (\bibinfo {year}
  {1998})%
  \bibAnnoteFile{NoStop}{Knill98}%
\bibitem{Jelezko06}%
  \BibitemOpen
  \bibfield{author}{%
  \bibinfo {author} {\bibfnamefont{F.}~\bibnamefont{Jelezko}}\ and\ \bibinfo
  {author} {\bibfnamefont{J.}~\bibnamefont{Wrachtrup}},\ }%
  \bibfield{journal}{%
  \Doi{10.1002/pssa.200671403}{\bibinfo {journal} {Physica Status Solidi (A)}}\
  }%
  \textbf{\bibinfo {volume} {203}},\ \bibinfo {pages} {3207} (\bibinfo {year}
  {2006})%
  \bibAnnoteFile{NoStop}{Jelezko06}%
\bibitem{Childress06}%
  \BibitemOpen
  \bibfield{author}{%
  \bibinfo {author} {\bibfnamefont{L.}~\bibnamefont{Childress}}, \bibinfo
  {author} {\bibfnamefont{M.~V.}\ \bibnamefont{Gurudev~Dutt}}, \bibinfo
  {author} {\bibfnamefont{J.~M.}\ \bibnamefont{Taylor}}, \bibinfo {author}
  {\bibfnamefont{A.~S.}\ \bibnamefont{Zibrov}}, \bibinfo {author}
  {\bibfnamefont{F.}~\bibnamefont{Jelezko}}, \bibnamefont{et~al.},\ }%
  \bibfield{journal}{%
  \Doi{10.1126/science.1131871}{\bibinfo {journal} {Science}}\ }%
  \textbf{\bibinfo {volume} {314}},\ \bibinfo {pages} {281} (\bibinfo {year}
  {2006})%
  \bibAnnoteFile{NoStop}{Childress06}%
\bibitem{Maze08}%
  \BibitemOpen
  \bibfield{author}{%
  \bibinfo {author} {\bibfnamefont{J.~R.}\ \bibnamefont{Maze}}, \bibinfo
  {author} {\bibfnamefont{P.~L.}\ \bibnamefont{Stanwix}}, \bibinfo {author}
  {\bibfnamefont{J.~S.}\ \bibnamefont{Hodges}}, \bibinfo {author}
  {\bibfnamefont{S.}~\bibnamefont{Hong}}, \bibinfo {author}
  {\bibfnamefont{J.~M.}\ \bibnamefont{Taylor}}, \bibnamefont{et~al.},\ }%
  \bibfield{journal}{%
  \Doi{10.1038/nature07279}{\bibinfo {journal} {Nature}}\ }%
  \textbf{\bibinfo {volume} {455}},\ \bibinfo {pages} {644} (\bibinfo {year}
  {2008})%
  \bibAnnoteFile{NoStop}{Maze08}%
\bibitem{Balasubramanian08}%
  \BibitemOpen
  \bibfield{author}{%
  \bibinfo {author} {\bibfnamefont{G.}~\bibnamefont{Balasubramanian}}, \bibinfo
  {author} {\bibfnamefont{I.-Y.}\ \bibnamefont{Chan}}, \bibinfo {author}
  {\bibfnamefont{R.}~\bibnamefont{Kolesov}}, \bibinfo {author}
  {\bibfnamefont{M.}~\bibnamefont{Al-Hmoud}}, \bibinfo {author}
  {\bibfnamefont{C.}~\bibnamefont{Shin}}, \bibnamefont{et~al.},\ }%
  \bibfield{journal}{%
  \Doi{10.1038/nature07278}{\bibinfo {journal} {Nature}}\ }%
  \textbf{\bibinfo {volume} {445}},\ \bibinfo {pages} {648} (\bibinfo {year}
  {2008})%
  \bibAnnoteFile{NoStop}{Balasubramanian08}%
\bibitem{Hanson08}%
  \BibitemOpen
  \bibfield{author}{%
  \bibinfo {author} {\bibfnamefont{R.}~\bibnamefont{Hanson}}, \bibinfo {author}
  {\bibfnamefont{V.~V.}\ \bibnamefont{Dobrovitski}}, \bibinfo {author}
  {\bibfnamefont{A.~E.}\ \bibnamefont{Feiguin}}, \bibinfo {author}
  {\bibfnamefont{O.}~\bibnamefont{Gywat}},\ and\ \bibinfo {author}
  {\bibfnamefont{D.~D.}\ \bibnamefont{Awschalom}},\ }%
  \bibfield{journal}{%
  \Doi{10.1126/science.1155400}{\bibinfo {journal} {Science}}\ }%
  \textbf{\bibinfo {volume} {320}},\ \bibinfo {pages} {352} (\bibinfo {year}
  {2008})%
  \bibAnnoteFile{NoStop}{Hanson08}%
\bibitem{Delange10}%
  \BibitemOpen
  \bibfield{author}{%
  \bibinfo {author} {\bibfnamefont{G.}~\bibnamefont{de~Lange}}, \bibinfo
  {author} {\bibfnamefont{Z.~H.}\ \bibnamefont{Wang}}, \bibinfo {author}
  {\bibfnamefont{D.}~\bibnamefont{RistÃ¨}}, \bibinfo {author}
  {\bibfnamefont{V.~V.}\ \bibnamefont{Dobrovitski}},\ and\ \bibinfo {author}
  {\bibfnamefont{R.}~\bibnamefont{Hanson}},\ }%
  \bibfield{journal}{%
  \Doi{10.1126/science.1192739}{\bibinfo {journal} {Science}}\ }%
  \textbf{\bibinfo {volume} {330}},\ \bibinfo {pages} {60} (\bibinfo {year}
  {2010})%
  \bibAnnoteFile{NoStop}{Delange10}%
\bibitem{Witzel06}%
  \BibitemOpen
  \bibfield{author}{%
  \bibinfo {author} {\bibfnamefont{W.~M.}\ \bibnamefont{Witzel}}\ and\ \bibinfo
  {author} {\bibfnamefont{S.~D.}\ \bibnamefont{Sarma}},\ }%
  \bibfield{journal}{%
  \Doi{10.1103/PhysRevB.74.035322}{\bibinfo {journal} {Phys. Rev. B}}\ }%
  \textbf{\bibinfo {volume} {74}},\ \bibinfo {eid} {035322} (\bibinfo {year}
  {2006})%
  \bibAnnoteFile{NoStop}{Witzel06}%
\bibitem{Rowan65}%
  \BibitemOpen
  \bibfield{author}{%
  \bibinfo {author} {\bibfnamefont{L.~G.}\ \bibnamefont{Rowan}}, \bibinfo
  {author} {\bibfnamefont{E.~L.}\ \bibnamefont{Hahn}},\ and\ \bibinfo {author}
  {\bibfnamefont{W.~B.}\ \bibnamefont{Mims}},\ }%
  \bibfield{journal}{%
  \Doi{10.1103/PhysRev.137.A61}{\bibinfo {journal} {Phys. Rev.}}\ }%
  \textbf{\bibinfo {volume} {137}},\ \bibinfo {pages} {A61} (\bibinfo {year}
  {1965})%
  \bibAnnoteFile{NoStop}{Rowan65}%
\bibitem{Haeberlen76}%
  \BibitemOpen
  \bibfield{author}{%
  \bibinfo {author} {\bibfnamefont{U.}~\bibnamefont{Haeberlen}},\ }%
  \emph{\bibinfo {title} {High Resolution NMR in Solids: Selective Averaging}}\
  (\bibinfo {publisher} {Academic Press Inc., New York},\ \bibinfo {year}
  {1976})%
  \bibAnnoteFile{NoStop}{Haeberlen76}%
\bibitem{Maze08b}%
  \BibitemOpen
  \bibfield{author}{%
  \bibinfo {author} {\bibfnamefont{J.~R.}\ \bibnamefont{Maze}}, \bibinfo
  {author} {\bibfnamefont{J.~M.}\ \bibnamefont{Taylor}},\ and\ \bibinfo
  {author} {\bibfnamefont{M.~D.}\ \bibnamefont{Lukin}},\ }%
  \bibfield{journal}{%
  \Doi{10.1103/PhysRevB.78.094303}{\bibinfo {journal} {Phys. Rev. B}}\ }%
  \textbf{\bibinfo {volume} {78}},\ \bibinfo {eid} {094303} (\bibinfo {year}
  {2008})%
  \bibAnnoteFile{NoStop}{Maze08b}%
\bibitem{Bollinger96}%
  \BibitemOpen
  \bibfield{author}{%
  \bibinfo {author} {\bibfnamefont{J.~J.}\ \bibnamefont{Bollinger}}, \bibinfo
  {author} {\bibfnamefont{W.~M.}\ \bibnamefont{Itano}}, \bibinfo {author}
  {\bibfnamefont{D.~J.}\ \bibnamefont{Wineland}},\ and\ \bibinfo {author}
  {\bibfnamefont{D.~J.}\ \bibnamefont{Heinzen}},\ }%
  \bibfield{journal}{%
  \Doi{10.1103/PhysRevA.54.R4649}{\bibinfo {journal} {Phys. Rev. A}}\ }%
  \textbf{\bibinfo {volume} {54}},\ \bibinfo {pages} {R4649} (\bibinfo {year}
  {1996})%
  \bibAnnoteFile{NoStop}{Bollinger96}%
\bibitem{Wineland94}%
  \BibitemOpen
  \bibfield{author}{%
  \bibinfo {author} {\bibfnamefont{D.~J.}\ \bibnamefont{Wineland}}, \bibinfo
  {author} {\bibfnamefont{J.~J.}\ \bibnamefont{Bollinger}}, \bibinfo {author}
  {\bibfnamefont{W.~M.}\ \bibnamefont{Itano}},\ and\ \bibinfo {author}
  {\bibfnamefont{D.~J.}\ \bibnamefont{Heinzen}},\ }%
  \bibfield{journal}{%
  \Doi{10.1103/PhysRevA.50.67}{\bibinfo {journal} {Phys. Rev. A}}\ }%
  \textbf{\bibinfo {volume} {50}},\ \bibinfo {pages} {67} (\bibinfo {year}
  {1994})%
  \bibAnnoteFile{NoStop}{Wineland94}%
\bibitem{Fischer09}%
  \BibitemOpen
  \bibfield{author}{%
  \bibinfo {author} {\bibfnamefont{J.}~\bibnamefont{Fischer}}, \bibinfo
  {author} {\bibfnamefont{M.}~\bibnamefont{Trif}}, \bibinfo {author}
  {\bibfnamefont{W.}~\bibnamefont{Coish}},\ and\ \bibinfo {author}
  {\bibfnamefont{D.}~\bibnamefont{Loss}},\ }%
  \bibfield{journal}{%
  \Doi{10.1016/j.ssc.2009.04.033}{\bibinfo {journal} {Solid State Comm.}}\ }%
  \textbf{\bibinfo {volume} {149}},\ \bibinfo {pages} {1443} (\bibinfo {year}
  {2009})%
  \bibAnnoteFile{NoStop}{Fischer09}%
\bibitem{Takahashi08}%
  \BibitemOpen
  \bibfield{author}{%
  \bibinfo {author} {\bibfnamefont{S.}~\bibnamefont{Takahashi}}, \bibinfo
  {author} {\bibfnamefont{R.}~\bibnamefont{Hanson}}, \bibinfo {author}
  {\bibfnamefont{J.}~\bibnamefont{van Tol}}, \bibinfo {author}
  {\bibfnamefont{M.~S.}\ \bibnamefont{Sherwin}},\ and\ \bibinfo {author}
  {\bibfnamefont{D.~D.}\ \bibnamefont{Awschalom}},\ }%
  \bibfield{journal}{%
  \Doi{10.1103/PhysRevLett.101.047601}{\bibinfo {journal} {Phys. Rev. Lett.}}\
  }%
  \textbf{\bibinfo {volume} {101}},\ \bibinfo {eid} {047601} (\bibinfo {year}
  {2008})%
  \bibAnnoteFile{NoStop}{Takahashi08}%
\bibitem{Huelga97}%
  \BibitemOpen
  \bibfield{author}{%
  \bibinfo {author} {\bibfnamefont{S.~F.}\ \bibnamefont{Huelga}}, \bibinfo
  {author} {\bibfnamefont{C.}~\bibnamefont{Macchiavello}}, \bibinfo {author}
  {\bibfnamefont{T.}~\bibnamefont{Pellizzari}}, \bibinfo {author}
  {\bibfnamefont{A.~K.}\ \bibnamefont{Ekert}}, \bibinfo {author}
  {\bibfnamefont{M.~B.}\ \bibnamefont{Plenio}}, \bibnamefont{et~al.},\ }%
  \bibfield{journal}{%
  \Doi{10.1103/PhysRevLett.79.3865}{\bibinfo {journal} {Phys. Rev. Lett.}}\ }%
  \textbf{\bibinfo {volume} {79}},\ \bibinfo {pages} {3865} (\bibinfo {year}
  {1997})%
  \bibAnnoteFile{NoStop}{Huelga97}%
\bibitem{Yang08}%
  \BibitemOpen
  \bibfield{author}{%
  \bibinfo {author} {\bibfnamefont{W.}~\bibnamefont{Yang}}\ and\ \bibinfo
  {author} {\bibfnamefont{R.-B.}\ \bibnamefont{Liu}},\ }%
  \bibfield{journal}{%
  \Doi{10.1103/PhysRevLett.101.180403}{\bibinfo {journal} {Phys. Rev. Lett.}}\
  }%
  \textbf{\bibinfo {volume} {101}},\ \bibinfo {eid} {180403} (\bibinfo {year}
  {2008})%
  \bibAnnoteFile{NoStop}{Yang08}%
\bibitem{Witzel10}%
  \BibitemOpen
  \bibfield{author}{%
  \bibinfo {author} {\bibfnamefont{W.~M.}\ \bibnamefont{Witzel}}, \bibinfo
  {author} {\bibfnamefont{M.~S.}\ \bibnamefont{Carroll}}, \bibinfo {author}
  {\bibfnamefont{A.}~\bibnamefont{Morello}}, \bibinfo {author}
  {\bibfnamefont{L.}~\bibnamefont{Cywi\ifmmode~\acute{n}\else \'{n}\fi{}ski}},\
  and\ \bibinfo {author} {\bibfnamefont{S.}~\bibnamefont{Das~Sarma}},\ }%
  \bibfield{journal}{%
  \Doi{10.1103/PhysRevLett.105.187602}{\bibinfo {journal} {Phys. Rev. Lett.}}\
  }%
  \textbf{\bibinfo {volume} {105}},\ \bibinfo {pages} {187602} (\bibinfo {year}
  {2010})%
  \bibAnnoteFile{NoStop}{Witzel10}%
\bibitem{Hahn50}%
  \BibitemOpen
  \bibfield{author}{%
  \bibinfo {author} {\bibfnamefont{E.~L.}\ \bibnamefont{Hahn}},\ }%
  \bibfield{journal}{%
  \Doi{10.1103/PhysRev.80.580}{\bibinfo {journal} {Phys. Rev.}}\ }%
  \textbf{\bibinfo {volume} {80}},\ \bibinfo {pages} {580} (\bibinfo {year}
  {1950})%
  \bibAnnoteFile{NoStop}{Hahn50}%
\bibitem{Waugh68}%
  \BibitemOpen
  \bibfield{author}{%
  \bibinfo {author} {\bibfnamefont{J.}~\bibnamefont{Waugh}}, \bibinfo {author}
  {\bibfnamefont{L.}~\bibnamefont{Huber}},\ and\ \bibinfo {author}
  {\bibfnamefont{U.}~\bibnamefont{Haeberlen}},\ }%
  \bibfield{journal}{%
  \Doi{10.1103/PhysRevLett.20.180}{\bibinfo {journal} {Phys. Rev. Lett.}}\ }%
  \textbf{\bibinfo {volume} {20}},\ \bibinfo {pages} {180} (\bibinfo {year}
  {1968})%
  \bibAnnoteFile{NoStop}{Waugh68}%
\bibitem{Burum79}%
  \BibitemOpen
  \bibfield{author}{%
  \bibinfo {author} {\bibfnamefont{D.~P.}\ \bibnamefont{Burum}}\ and\ \bibinfo
  {author} {\bibfnamefont{W.~K.}\ \bibnamefont{Rhim}},\ }%
  \bibfield{journal}{%
  \Doi{10.1063/1.438385}{\bibinfo {journal} {J. Chem. Phys.}}\ }%
  \textbf{\bibinfo {volume} {71}},\ \bibinfo {pages} {944} (\bibinfo {year}
  {1979})%
  \bibAnnoteFile{NoStop}{Burum79}%
\bibitem{Carr54}%
  \BibitemOpen
  \bibfield{author}{%
  \bibinfo {author} {\bibfnamefont{H.~Y.}\ \bibnamefont{Carr}}\ and\ \bibinfo
  {author} {\bibfnamefont{E.~M.}\ \bibnamefont{Purcell}},\ }%
  \bibfield{journal}{%
  \Doi{10.1103/PhysRev.94.630}{\bibinfo {journal} {Phys. Rev.}}\ }%
  \textbf{\bibinfo {volume} {94}},\ \bibinfo {pages} {630} (\bibinfo {year}
  {1954})%
  \bibAnnoteFile{NoStop}{Carr54}%
\bibitem{Meiboom58}%
  \BibitemOpen
  \bibfield{author}{%
  \bibinfo {author} {\bibfnamefont{S.}~\bibnamefont{Meiboom}}\ and\ \bibinfo
  {author} {\bibfnamefont{D.}~\bibnamefont{Gill}},\ }%
  \bibfield{journal}{%
  \bibinfo {journal} {Rev. Sc. Instr.}\ }%
  \textbf{\bibinfo {volume} {29}},\ \bibinfo {pages} {688} (\bibinfo {year}
  {1958})%
  \bibAnnoteFile{NoStop}{Meiboom58}%
\bibitem{Hartman62}%
  \BibitemOpen
  \bibfield{author}{%
  \bibinfo {author} {\bibfnamefont{S.~R.}\ \bibnamefont{Hartman}}\ and\
  \bibinfo {author} {\bibfnamefont{E.~L.}\ \bibnamefont{Hahn}},\ }%
  \bibfield{journal}{%
  \Doi{10.1103/PhysRev.128.2042}{\bibinfo {journal} {Phys. Rev.}}\ }%
  \textbf{\bibinfo {volume} {128}},\ \bibinfo {pages} {2042} (\bibinfo {year}
  {1962})%
  \bibAnnoteFile{NoStop}{Hartman62}%
\bibitem{Weis06}%
  \BibitemOpen
  \bibfield{author}{%
  \bibinfo {author} {\bibfnamefont{V.}~\bibnamefont{Weis}}\ and\ \bibinfo
  {author} {\bibfnamefont{R.}~\bibnamefont{Griffin}},\ }%
  \bibfield{journal}{%
  \Doi{DOI: 10.1016/j.ssnmr.2005.08.005}{\bibinfo {journal} {Solid State NMR}}\
  }%
  \textbf{\bibinfo {volume} {29}},\ \bibinfo {pages} {66 } (\bibinfo {year}
  {2006})%
  \bibAnnoteFile{NoStop}{Weis06}%
\bibitem{Khutsishvili66}%
  \BibitemOpen
  \bibfield{author}{%
  \bibinfo {author} {\bibfnamefont{G.~R.}\ \bibnamefont{Khutsishvili}},\ }%
  \bibfield{journal}{%
  \bibinfo {journal} {Sov. Phys. Uspekhi}\ }%
  \textbf{\bibinfo {volume} {8}},\ \bibinfo {pages} {743} (\bibinfo {year}
  {1966})%
  \bibAnnoteFile{NoStop}{Khutsishvili66}%
\bibitem{Ramanathan08}%
  \BibitemOpen
  \bibfield{author}{%
  \bibinfo {author} {\bibfnamefont{C.}~\bibnamefont{Ramanathan}},\ }%
  \bibfield{journal}{%
  \Doi{10.1007/s00723-008-0123-7}{\bibinfo {journal} {App. Mag. Res.}}\ }%
  \textbf{\bibinfo {volume} {34}},\ \bibinfo {pages} {409} (\bibinfo {year}
  {2008})%
  \bibAnnoteFile{NoStop}{Ramanathan08}%
\bibitem{Togan11}%
  \BibitemOpen
  \bibfield{author}{%
  \bibinfo {author} {\bibfnamefont{E.}~\bibnamefont{Togan}}, \bibinfo {author}
  {\bibfnamefont{Y.}~\bibnamefont{Chu}}, \bibinfo {author}
  {\bibfnamefont{A.}~\bibnamefont{Imamoglu}},\ and\ \bibinfo {author}
  {\bibfnamefont{M.~D.}\ \bibnamefont{Lukin}},\ }%
  \bibfield{journal}{%
  \Doi{10.1038/nature10528}{\bibinfo {journal} {Nature}}\ }%
  \textbf{\bibinfo {volume} {478}},\ \bibinfo {pages} {497} (\bibinfo {year}
  {2011})%
  \bibAnnoteFile{NoStop}{Togan11}%
\bibitem{Meriles10}%
  \BibitemOpen
  \bibfield{author}{%
  \bibinfo {author} {\bibfnamefont{C.~A.}\ \bibnamefont{Meriles}}, \bibinfo
  {author} {\bibfnamefont{L.}~\bibnamefont{Jiang}}, \bibinfo {author}
  {\bibfnamefont{G.}~\bibnamefont{Goldstein}}, \bibinfo {author}
  {\bibfnamefont{J.~S.}\ \bibnamefont{Hodges}}, \bibinfo {author}
  {\bibfnamefont{J.}~\bibnamefont{Maze}}, \bibnamefont{et~al.},\ }%
  \bibfield{journal}{%
  \Doi{10.1063/1.3483676}{\bibinfo {journal} {J. Chem. Phys.}}\ }%
  \textbf{\bibinfo {volume} {133}},\ \bibinfo {pages} {124105} (\bibinfo {year}
  {2010})%
  \bibAnnoteFile{NoStop}{Meriles10}%
\end{thebibliography}

\end{document}